# Luminescence Properties of Individual Empty and Water-filled Single-Walled Carbon Nanotubes

*Sofie Cambré[1,2,3], Silvia M. Santos[2,3], Wim Wenseleers[1,*], Ahmad R. T. Nugraha[4], Riichiro Saito[4], Laurent Cognet[2,3], Brahim Lounis[2,3,*]*

[1] Experimental Condensed Matter Physics Laboratory, University of Antwerp, Belgium

[2] Univ Bordeaux, LP2N, F-33405 Talence, France

[3] CNRS & Institut d'Optique, LP2N, F-33405 Talence, France

[4] Department of Physics, Tohoku University, Sendai 980-8578, Japan

Correspondence: wim.wenseleers@ua.ac.be; blounis@u-bordeaux1.fr

**RECEIVED DATE (to be automatically inserted after your manuscript is accepted if required according to the journal that you are submitting your paper to)**

ABSTRACT

The influence of water-filling on the photoluminescence (PL) properties of SWCNTs is studied by ensemble and single-molecule PL spectroscopy. Red-shifted PL and PL excitation spectra are observed upon water filling for 16 chiralities and can be used to unambiguously distinguish empty SWCNTs from filled ones. The effect of water-filling on the optical transitions is well reproduced by a continuum dielectric constant model previously developed to describe the influence of the nanotube outer environment. Empty nanotubes display narrower luminescence lines and lower inhomogeneous broadening, signatures of reduced extrinsic perturbations. The radial breathing mode (RBM) phonon side band is clearly observed in the PL spectrum of small diameter empty tubes and a strong exciton-phonon coupling is measured for this vibration. Biexponential PL decays are observed for empty and water filled tubes, and only the short living component is influenced by the water-filling. This may be attributed to a shortening of the radiative lifetime of the bright state by the inner dielectric environment.

Single-walled carbon nanotubes (SWCNTs) have unique optical properties[1] due to their true one-dimensional nature. Because of the confinement, strong Coulomb interactions between electrons and holes lead to the formation of excitons,[2-4] with large binding energies that can reach up to ~1eV in air-suspended semiconducting nanotubes.[5] When nanotubes are contained in a dense medium, dielectric screening lowers the energy of the excitons, leading to red-shifts of the optical transitions.[5-7] Since all atoms of a SWCNT are at its surface, the photoluminescence (PL) properties of SWCNTs are extremely sensitive to the local environment[8, 9] and to the presence of structural defects that can severely quench the PL.[10] While the effect of the nanotubes outer environment on their optical properties has been widely studied, little is known on the influence of their inner content.

Aqueous surfactant solutions of chemically purified tubes or solutions of ultrasonicated tubes mainly contain water-filled tubes.[11] Since heavy ultrasonication steps are almost universally used to prepare solutions for optical characterization, nearly all spectroscopic studies of aqueous solutions reported to date are based on water-filled tubes.[12]

Raman studies have shown that water-filling leads to a blue-shift of the radial breathing mode (RBM) vibrations of the nanotubes attributed to an additional restoring force for the RBM vibration excerted by the encapsulated water.[11, 13] In addition, a global red-shift of the electronic transitions observed by two dimensional (2D) resonant Raman scattering in ensemble experiments was a first indication of reduced dielectric screening in empty nanotubes.[11, 13] Apart from those Raman studies and preliminary fluorescence experiments,[14] there is still a patent lack of knowledge about the spectroscopic differences between empty and water-filled SWCNTs.

In this work we combine ensemble and single-molecule experiments to study the effect of the water-filling on the PL and PL-excitation (PLE) spectra of SWCNTs. Clear shifts of the optical transitions are observed and studied for a wide range of chiralities. The knowledge of the effect of water-filling on the electronic transitions is crucial to understand *e.g.* the different $E_{ii}$ values that have been reported in the literature. The effect of water-filling was also studied spatially along the length of the tube, yielding for example information on the degree of filling and the presence of defects. The results show that empty nanotubes are model systems to study the intrinsic behavior of pristine nanotubes in liquid environments.

Throughout this study, we use bile salt surfactants to solubilize the SWCNTs in aqueous solutions,[15] in order to individualize the nanotubes, without using ultrasonication, which is known to cause structural damage to the SWCNTs.[11] Density gradient ultracentrifugation (DGU) was used to separate empty and filled HiPco nanotubes,[14] Fractions containing exclusively empty nanotubes and fractions containing a majority of filled tubes with a small proportion of empty nanotubes (less than 10%) were obtained. The ratio of empty/filled tubes in the different fractions was obtained by resonant Raman scattering (see Supporting Information Figures S1-S2).

Figure 1a shows an overlay of 2D PLE maps measured on empty (red) and filled (blue) SWCNTs fractions. Although $E_{11}$ (emission) and $E_{22}$ (excitation) optical transitions of the two fractions overlap, they display distinct maxima. As well, a significant line broadening (~ 10 to 30%) is observed for water-filled tubes (Figure 1a and Supporting

Information Figures S3-S10). To provide a quantitative determination of the shifts ($\Delta E_{11}$ and $\Delta E_{22}$) for each chirality, a global fit of samples containing different ratios of empty/filled tubes was performed. More precisely, $E_{11}$ and $E_{22}$ transition energies are obtained by fitting 1D PL and PLE spectra such as those shown in the top and right panels of Figure 1a (see also Supporting Information Figures S7-S10). The obtained values of $\Delta E_{11}$ and $\Delta E_{22}$ are presented on Figure 1b showing that water-filling leads to red-shifts, depending on the specific nanotube diameter, ranging from 5 to 30 meV as expected for an increased dielectric screening. We note that PLE spectra are systematically broader and more asymmetric than PL emission lines[16] explaining the larger error bars obtained for $E_{22}$ shifts as compared to $E_{11}$.

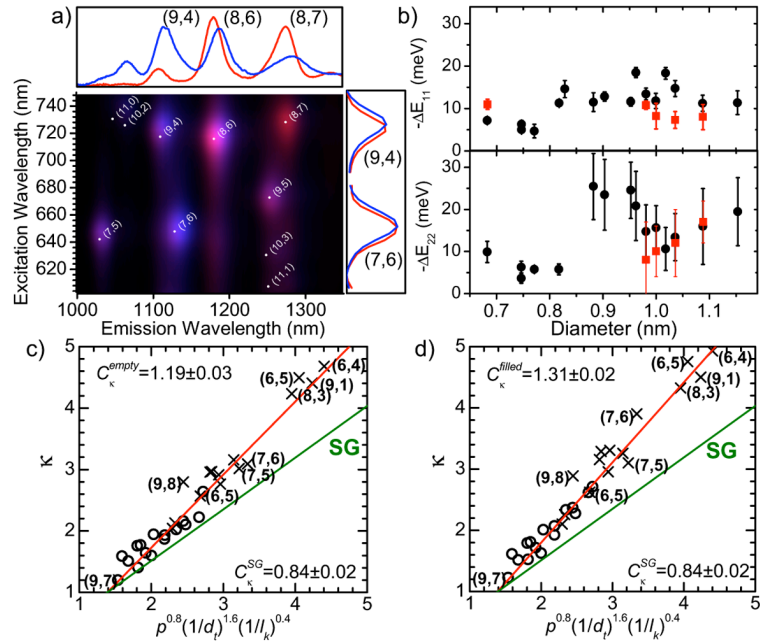

***Figure 1:*** *a) Two-color overlay graph of 2D PLE maps of a DGU sorted sample of empty tubes (red) and one containing mainly water-filled SWCNTs (blue); (see also Figures S3-S6 for other 2D PLE maps). Top and right panels: PL (integrated over excitation wavelength range 700-745nm) and PLE (integrated over emission wavelength range 1100-1150nm) spectra for empty*

*(red) and water-filled (blue) SWCNTs. b) Electronic red-shifts of the first ($\Delta E_{11}$) and second ($\Delta E_{22}$) optical transitions as a function of nanotube diameter[17] (see also table S1 in the Supporting Information) obtained from ensemble (black balls) and single-molecule (red squares) experiments. c-d) effective dielectric constant $\kappa$ reproducing the experimental $E_{ii}$ values for empty (c) and water-filled (d) SWCNTs (see text). Circle (cross) symbols are for $E_{11}$ ($E_{22}$) transitions. Red lines correspond to the linear fit of the data, yielding the slope $C_\kappa$ for each sample. The green line corresponds to the fit of the reference supergrowth (SG) sample used in [6].*

A model to account for the dielectric screening of the external environment was previously proposed by Nugraha *et al*. [6]. Within this model an effective dielectric constant $\kappa$ is derived for each experimental $E_{ii}$ value, and plotted as a function of $\kappa = C_\kappa p^{0.8}(1/d_t)^{1.6}(1/l_k)^{0.4}$, where $p$ is the subband index ($p$=1,2 for $E_{11}$ and $E_{22}$ respectively), $d_t$ the tube diameter and $l_k$ the exciton size in the reciprocal space, in which ($1/d_t$) and ($1/l_k$) are set to be dimensionless values.[6] The linear slope, $C_\kappa$, observed in such a plot yields a measure of the dielectric screening of the nanotubes (both semiconducting and metallic), in different environments. For these fits, the PL line widths are taken as error bars on the determination of $\kappa$ in order to account for inhomogeneous broadening effects. Using the same unit definitions as in reference [6], Figures 1c-d gives these plots for empty and water-filled SWCNTs as well as for $C_\kappa^{SG}$=0.84±0.02 corresponding to the reference sample of as-produced 'Super Growth' nanotubes (the free-standing nanotube forests used in ref. [18], which are generally considered as the best approximation to isolated nanotubes without dielectric screening). Water-filling leads to a significant increase of $C_\kappa$ compared to empty and SG nanotubes, showing that $C_\kappa$ can account for both the outer *and* inner environment. This continuum model cannot however reproduce deviations from the linear slope in Figs. 1c-d, which could originate from subtle variations of $\Delta E_{11}$ ($\Delta E_{22}$) with the ordering of the water-molecules in nanotubes of different diameters, and would rather require a molecular model.

To overcome the effects of inhomogeneous broadening and averaging inherent to ensemble measurements we performed single molecule experiments. These experiments give access to distributions of physical parameters (instead of their average) and allow cross-correlations between distributions to be determined. Furthermore, they also provide the possibility to follow the spatial changes of the parameters along the length of the nanotube.

Individual nanotubes with different diameters and belonging to different chiralities were studied (chiral indices (12,1), (11,3), (10,5) and (9,7) of the same family $2n+m=25$ and the smaller diameter (6,4) nanotubes. In order to avoid any bias introduced by the length of the nanotubes (tube-ends effect) or the number of defects present on the tube, we studied, for each chirality, a large number of tubes with various lengths, ranging from the diffraction limit ($\sim$500nm) up to 10 microns, and various PL intensities.

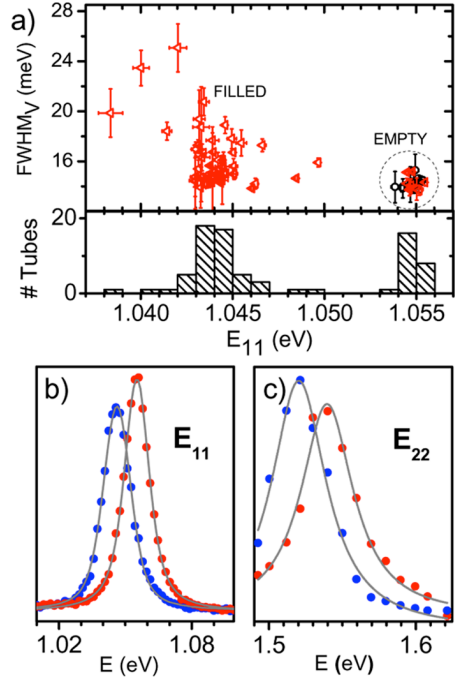

***Figure 2****: a) linewidths (FWHM$_V$) versus E$_{11}$ obtained by fitting with Voigt profiles the PL spectra of 78 different (12,1) tubes selected from DGU sorted empty nanotube samples (black circles) and samples containing both empty and filled tubes (red triangles). Tubes assigned as empty are found in the dotted circle. Lower panel: histogram of the E$_{11}$ peak emission energies. Comparative PL (b) and PLE (c) spectra of empty (red circles) and water-filled (blue circles) individual (12,1) tubes and their fits (gray curves).*

Figure 2b shows the examples of PL spectra obtained from two (12,1) nanotubes, a water filled and an empty one. Emission peaks of water-filled tubes are significantly red-shifted in agreement with ensemble measurements (red data points in Figure 1b). We use Voigt line shapes (see below and Supporting Information Figures S11-S12 for justification of the shape) to fit the spectra and to extract the peak positions ($E_{11}$) and the width ($FWHM_V$) of the emission lines. In Figure 2a we present the cross-correlation between $E_{11}$ and $FWHM_V$ measured on 78 individual (12,1) tubes. The empty nanotubes (extracted from the DGU sorted sample) display a narrow distribution of emission peaks centered on 1.055eV (close to the peak emissions of the bulk measurement) and a narrow distribution of $FWHM_V$ centered on ~14meV a value which is, as expected, smaller than the width of the inhomogeneously-broadened bulk emission line. In comparison, nanotubes extracted from the fractions containing a majority of filled tubes, show a more heterogeneous distribution. Since this fraction contains a small proportion of empty ones, some few emission lines fall within the distributions just described above (dashed circle) while the majority is centered at a different mean and with a larger spread. The same behavior is obtained for the (11,3), (10,5) and (9,7) tubes as well as for the thinner nanotubes (6,4) (Figures S13-S14, Supporting Information). This indicates that water filling leads to a broadening of the PL lines (up to ~20% in average, depending on the chirality) and to wider distributions of both $E_{11}$ and $FWHM_V$. The strong broadening observed in the ensemble experiments for water-filled tubes is thus originating from a combination of larger standard deviations (inhomogeneous broadening), and larger line widths (homogeneous broadening) (see below for more detailed description of the line shape).

We next studied the second optical transitions by measuring PLE spectra at the single nanotube level. For this purpose, a tunable Ti:Saphire laser was used for resonantly exciting the nanotubes at their $E_{22}$ (Figure 2c). Similarly to $E_{11}$ transitions, the $E_{22}$ optical transitions are red-shifted, broadened and display a larger statistical distribution for the water-filled tubes (Supporting Figure S12) revealing again the presence of both homogeneous and inhomogeneous effects in ensemble experiments. All together, these results indicate that nanotubes can securely be assigned at the single molecule level as empty or water-filled tubes through their luminescence lines, as is the case for Raman spectra.

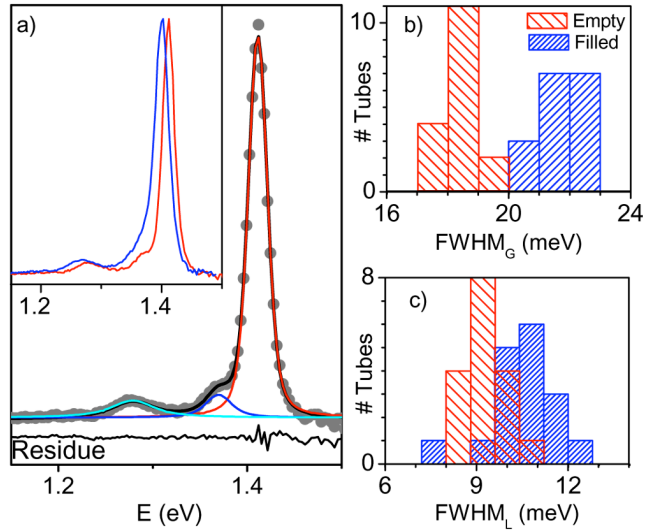

***Figure 3:*** *a) PL spectrum of an empty (6,4) tube (gray dots) fitted (black) using a superposition of the main $E_{11}$ excitonic peak (red), the RBM phonon side band (blue) and the phonon side band of the K-momentum dark exciton (cyan). Inset: comparison of the PL-spectra of an empty (6,4) tube (red) with that of a water-filled one (blue). b-c) Histograms of the Gaussian ($FWHM_G$) and Lorentzian ($FWHM_L$) line-width contributions of the Voigt fits of the main $E_{11}$ excitonic peak for empty (red) and water-filled (blue) (6,4) SWCNTs.*

We now focus on the PL spectral shape of small diameter empty (6,4) tubes. Besides the main excitonic peak and the K momentum dark-exciton phonon sideband at ~130 meV,[19-22] an additional sideband is clearly resolved which can be assigned to the RBM phonon sideband, positioned at $\hbar\omega_{RBM}$ ~41.5 meV for these thin tubes (Figure 3a). The clear observation of the RBM phonon sideband for empty (6,4) tubes suggest that it is responsible for the asymmetric line shape observed for the filled (6,4) tubes (inset of Figure 3a) and for the other small-diameter nanotubes. Up to now, the RBM sideband in the PL spectrum had only been resolved in low temperature PLE single molecule experiments[23] and also evoked to explain asymmetric $E_{11}$ lines of room temperature ensemble PLE spectra.[24] Our single tube PL spectra are well fitted with three Voigt curves (see Supporting information Figure S12). The Voigt line shapes include both the intrinsic line broadening (Lorentzian) due to the dipole dephazing time as well as the inhomogeneous broadening (Gaussian) due to exciton transition energy fluctuations induced by environmental heterogeneities. In the fits, the RBM sideband position was fixed at an energy of 41.5 meV below $E_{11}$, a value accurately determined by Raman spectroscopy for empty tubes[13] and the Gaussian contribution of all three Voigt lines had identical widths. The relative weight of the RBM phonon sideband to the $E_{11}$ main emission peak defines the Huang-Rhys factor S= $V^2_{RBM}/\hbar^2\omega^2_{RBM}$ with $V_{RBM}$, the exciton-phonon coupling energy. We find S~0.1 which leads to $V_{RBM}$ ~ 13meV, a value larger than reported previously (4-5 meV) for the $E_{22}$-transition in ~1nm diameter tubes.[25] This is consistent with theoretical predictions of a stronger exciton-phonon coupling for small diameter tubes.[26-28] Apart from the red-shift of the peak emission, water-filled nanotubes also display wider emission lines. Interestingly, Figure 3b-c shows that both the Lorentzian and the Gaussian components are broadened suggesting a faster exciton dephasing time and increased environmental fluctuations in water filled tubes. The dephazing times extracted from the Lorentzian component width (~9 meV) on single nanotubes is in agreement with that measured using ultrafast pump probe spectroscopy on ensembles of tubes (~7 to 8 meV for (6,5) tubes).[29, 30]

We now address the influence of water-filling on the recombination dynamics of excitons in SWCNTs. We have previously shown that several extrinsic factors affect their PL decays. Indeed, (6,5) and (6,4) nanotubes can exhibit either mono- or biexponential behaviors depending on the synthesis methods and their environments.[9, 31] These observations were explained using a model which takes into account the band edge exciton fine structure[1] (including the two lowest bright and dark singlet states) and the dominant defect-dependent non-radiative decay mechanisms proposed by Pereibenos *et al*.[32] More precisely, individual high-quality tubes display a biexponential PL decay with two components, a short time component ($\tau_S$, of the order of 50ps, reflecting essentially the decay of the bright state) and a long one ($\tau_L$, up to a few ns, reflecting the decay of the dark one, lying a few meV below the bright one). In contrast, in nanotubes subject to defects or environmental effects, an extrinsic fast non-radiative relaxation process dominates $\tau_L$ and shortens $\tau_S$ leading to a mono-exponential decay.[9]

Figure 4a shows a typical PL decay curve obtained from an empty (6,4) tube. A biexponential decay is observed with time constants $\tau_S$ ~70ps and $\tau_L$ ~1.6ns, and long time component fractional yield ($A_{long}$) of 7%. Figure 4b compares $\tau_S$ and $\tau_L$ for individual empty and filled (6,4) tubes. Remarkably, $\tau_L$ (and $A_{long}$ data not shown) is not significantly influenced by the water-filling, while $\tau_S$ is shorter for the filled tubes. Although one cannot absolutely rule out a mechanism that would increase the non-radiative recombination of the bright exciton solely, we believe that this behavior is not primarily due to modifications of the magnitude of non-radiative decay pathways, since one expects that such modifications would drastically affect the long time component of the PL decay. Indeed, it has been shown both experimentally[9] and theoretically[32] that extrinsic effects which drive non-radiative decay processes affect all band-edge excitonic singlet sublevels. Therefore, we propose that the sole reduction of $\tau_S$ upon water-filling may be due to an increased exciton radiative decay rate as a consequence of a higher dielectric constant inside the tubes.[33]

We found variation of the decay rates ($1/\tau_S$) between empty and filled of the order of ~$1 ns^{-1}$, a value which is of the same order of the calculated radiative decays in SWCNTs.[34, 35] The modification of the radiative lifetime of a dipole placed at the interface between two media in a nanotube geometry is complicated to model, and strong variations are predicted for planar interface.[36]

We finally focus on the variations of PL properties along the length of tubes. Interestingly, PL spectra recorded at different positions of long empty and filled nanotubes ($>5\mu$m) did not show any significant variations of $E_{11}$ and $FWHM_V$. This indicates that tubes are either completely empty or completely filled and do not present distinct empty and water-filled domains within a same tube, in agreement with earlier DGU separation experiments.[14, 37] However, biexponential decays with the largest $A_{long}$ and longest $\tau_L$ are only observed around the center of the tubes, regardless of water filling. Indeed, as shown in Figure 4c, $A_{long}$ decreases towards the tube ends and vanishes completely at the extremities. This shows that the tube ends act as defect centers which introduce additional non-radiative decay channels responsible for the observed fast mono-exponential PL decays. This also explains why biexponential decays are only observed for long (>1$\mu$m) SWCNTs.

We note that in previous studies of SWCNT PL decays,[9] biexponential behaviors were rarely observed for (6,4) tubes (as opposed to (6,5) tubes). Here, we systematically observe biexponential decays both for empty and water-filled (6,4) SWCNTs, provided that care is taken of the following factors: (1) highly diluted samples are used to minimize energy transfer to other tubes (towards this aim, DGU-separated solutions were valuable) (2) non-sonicated solutions are used to obtain long, bright and nearly defect-free SWCNTs and (3) the center of long tubes is studied and not the tube ends.

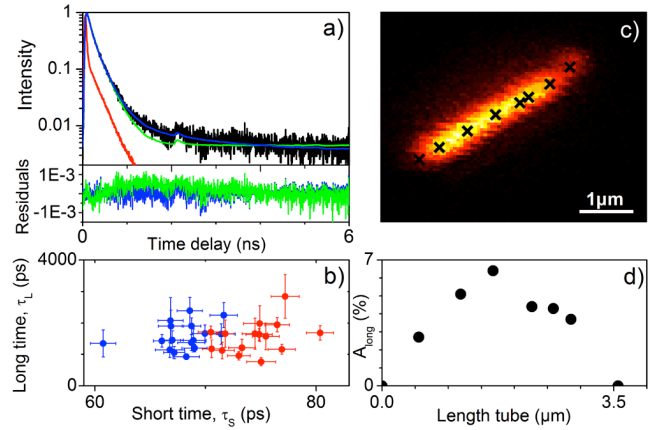

***Figure 4***: *a) PL decay of a filled (6,4) SWCNT (black) and its fits to a bi-exponential decay (blue) as the mono-exponential fit (green) does not reproduce the experimental data. The instrumental response function is shown in red. Bottom panel gives the residuals for mono (green) and biexponential (blue) fits. b) Long versus short decay time components for empty (black) and filled (red) (6,4) SWCNTs. c) Positions at which PL decay measurements were performed along an empty (6,4) tube superimposed with a*

*confocal PL image of the tube.. Inset: fractional intensity ($A_{long}$) of the long-time component, $\tau_L$, of the biexponential fit along the length of this tube. At the tube ends, monoexpoential decays are observed ($A_{long}$ =0).*

CONCLUSIONS

In this work, we combine ensemble and single-molecule experiments to study the effect of water-filling on the PL properties of DGU empty/filled sorted SWCNTs. PL and PLE spectra show red shifts of the optical transitions upon water filling, which can be used to unambiguously distinguish empty SWCNTs from filled ones. Empty individual nanotubes have reduced extrinsic perturbations, compared to the more commonly encountered water-filled ones, resulting in narrow emission lines with well-resolved RBM phonon side bands for the small diameter (6,4) tubes. Time resolved PL studies show that only the short living component is influenced by the water-filling, presumably due to a shortening of the radiative lifetime of the bright state by the inner dielectric environment.

METHODS Raw HiPco SWCNTs were solubilized in $D_2O$ with the surfactant sodium deoxycholate (DOC, 99% Acros organics), using only (gentle) magnetic stirring as described in ref. [15] (no ultrasonication is applied). The empty tubes were then sorted out using density gradient ultracentrifugation as described in ref. [14] with a 0.7%w/v surfactant concentration, and were characterized with absorption, resonant Raman scattering and 2D PLE spectroscopy to obtain the composition of the sample (see Supporting Information). Several batches of HiPco samples were used depending on the abundance of the chiralities to be studied.[38]

2D PLE spectroscopy was performed using a home built set-up based on either a liquid-nitrogen cooled extended InGaAs photodiode array detector (Roper Scientific, OMA V: 1024/LN-2.2, sensitive up to 2.2 $\mu$m) or a liquid-nitrogen cooled deep depletion Si CCD (Roper Scientific, SPEC-10:400BR, sensitive up to ~1$\mu$m) and a pulsed Xe-lamp for excitation (Edinburgh Instruments, Xe900/xP920).

For the single-molecule experiments the SWCNTs were immobilized in aqueous agarose gels (5 wt%), which do not change the properties of the solubilized nanotubes compared to aqueous solutions.[8, 10] Single-molecule wide-field and confocal photoluminescence microscopes were used to image individual SWCNTs. The SWCNTs were optically excited with a CW or pulsed laser near their second order resonance $E_{22}$.[21] For this purpose we used a 561nm solid state laser and a tunable Ti:sapphire laser, and for the pulsed excitation an optical parametric oscillator (OPO; operated at 582nm, 76 MHz repetition rate, ~2ps pulse width). For the PL decays a conventional time-correlated single-photon counting setup was used. Excitation powers (<31W/cm$^2$) were kept low enough in order to assure that less than one photon is absorbed per pulse, per micron tube length, ensuring that multi-excitonic effects are absent.

ACKNOWLEDGMENT: We thank S. K. Doorn and J. G. Duque for helpful discussion. This work was funded by the Agence Nationale de la Recherche, Région Aquitaine, the French ministry of education and research, and the European Research Council. S. C. and W. W. gratefully acknowledge the financial support from the Fund for Scientific Research Flanders, Belgium (FWO-Vlaanderen: projects G.0129.07, G.0400.11 and G.0211.12), which also provided S.C. a postdoctoral fellowship and a mobility grant for visiting the Bordeaux group. RS acknowledges MEXT grant No 20241023.

**Supporting Information Available:** DGU separation of empty and filled tubes, 2D bulk PL-EX fits and single molecule statistics for other tube chiralities. This material is available free of charge *via* the internet at http://pubs.acs.org.

TOC figure

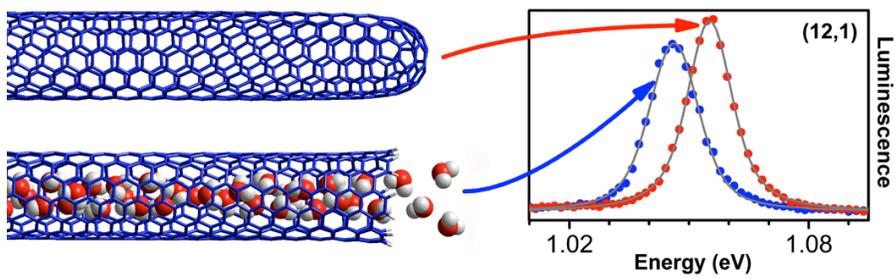